\documentclass[pra,twocolumn, showpacs,showkeys,superscriptaddress,amsmath,amssymb]{revtex4-1}
\usepackage{graphicx}
\usepackage{color}
\usepackage{amssymb,amsmath}
\usepackage{comment}

\begin{document}

\title{The Bose Hubbard model with squeezed dissipation}

\author{Fernando Quijandr\'{\i}a} 

\affiliation{Instituto de Ciencia de Materiales de Arag\'on y
  Departamento de F\'{\i}sica de la Materia Condensada,
  CSIC-Universidad de Zaragoza, Zaragoza, E-50012, Spain.}

\author{Uta Naether} 

\affiliation{Instituto de Ciencia de Materiales de Arag\'on y
  Departamento de F\'{\i}sica de la Materia Condensada,
  CSIC-Universidad de Zaragoza, Zaragoza, E-50012, Spain.}


\author{Diego Porras}
\affiliation{Department of Physics and Astronomy, University of Sussex, Brighton BN1 9QH, United Kingdom}

\author{Juan Jos\'e Garc\'{\i}a-Ripoll}
\affiliation{Instituto de F\'{\i}sica Fundamental, IFF-CSIC, Serrano 113-bis, Madrid E-28006, Spain}

\author{David Zueco} 
\affiliation{Instituto de Ciencia de Materiales de Arag\'on y
  Departamento de F\'{\i}sica de la Materia Condensada,
  CSIC-Universidad de Zaragoza, Zaragoza, E-50012, Spain.}
\affiliation{Fundaci\'on ARAID, Paseo Mar\'{\i}a Agust\'{\i}n 36, Zaragoza 50004, Spain}

\date{\today}

\begin{abstract}
The stationary properties of the Bose-Hubbard model under squeezed
dissipation are investigated.  
The dissipative model does not possess a $U(1)$ symmetry, but parity
is conserved: $\langle a_j \rangle \to -\langle a_j \rangle$.
We find that
$\langle a_j \rangle = 0$ always holds, so no symmetry breaking occurs.
Without the onsite repulsion, the linear case is known to be
critical.
At the critical point the system freezes to an EPR state with infinite
two mode entanglement.
We show  here that the correlations are rapidly destroyed whenever the
repulsion is switched on.
Then, the system approaches a thermal state with 
an effective temperature defined in terms of the squeezing parameter in the dissipators.
We characterize this transition by means of a Gutzwiller {\it ansatz}
and the Gaussian Hartree-Fock-Bogoliubov approximation.
\end{abstract}

\pacs{Valid PACS appear here}
\maketitle



\section{Introduction}

Whenever a system is coupled 
to a large and uncontrollable environment, effective irreversibility arises  \cite{Breuer2002, Weiss2008,
  Rivas2011}.  The environment can
be anything except the system of interest. Common examples are  the electromagnetic
radiation, the gravitational field, the phonons or the electrical noise.
Typically, the interaction with the bath 
provides equilibration, {\it i.e.} the
appearance of a stationary state.
Typical is also that, in macroscopic systems, the coupling with the outside is
through the surface.  The ratio surface/volume, being finite for
providing the irreversibility and the
fluctuation-dissipation relation, is sufficiently small to be neglected  in the
equilibrium state.  Thus, the equilibrium density matrix is given by
{\it Gibbs}, $\varrho^* \sim {\rm e}^{-\beta H_S}$. Here $H_s$ is the
system Hamiltonian, meaning that the interaction part of the total
Hamiltonian is neglected \cite{Balescu1975}.  
The above paradigm, with an unquestionable success, starts to fail as
soon as the interaction part ($H_I$) is no longer a perturbation over the bulk
($H_S$) and  {\it Gibbs} is not  the stationary solution.
Both system and interaction contribute to account for the equilibrium
properties \cite{Hanggi2005, Pachon2014}.  

An arena for dealing with such a situation  are man-made
realizations of few level systems, as qubits (two level systems). 
Examples could be superconducting circuits, ion traps  or
quantum dots.  Though they behave as few level systems,  they are  macroscopic due to their 
coupling to the environment.  
Maybe motivated by this mesoscopic physics
there is a theoretical literature trying to characterize the equilibrium
properties of systems {\it driven by dissipation} ( as usually termed
through the papers).
Roughly, the equilibrium statistical mechanics is now extended
considering the bath and the type of system-environment interaction.  
This  {\it extra} dependence comes with  some richness on the
equilibrium states and their phases \cite{Tomadin2011, Tomadin2012,
  Horstmann2012, Diehl2010, Eisert2010, Jin2013, Ruiz-Rivas2014,
  Jin2014, LeBoite2013,
  Grujic2013, Boite2014}. Besides, it is also found that
this dependence provides an extra way of control for quantum states.
For example, environment engineering can be used for state preparation
\cite{Torre2012, Quijandria2013}. 

A paradigm in many-body physics is the
Bose-Hubbard (BH) model.  It appears in many  different contexts,
it has been realized experimentally  and solved within all the approximations
developed \cite{Cazalilla2011}.
One more study of the BH is the one presented here.  We study the
equilibrium properties of the model whenever squeezed dissipation is
taken into account.  Squeezed noise provides
long-range correlations, producing even a critical point in Gaussian
models \cite{Eisert2010, Quijandria2013}.   This long order correlation
competes with the self-interaction of the model (characterized in this
work by the strength $U$).  The phenomenology  that we find
is rather simple. We extensively study the model, finding that the
model does not condensate, thus
$\langle a_j \rangle = 0$ always.  Both the self-interaction and the squeezing
competes, and the system become critical if $U=0$. At the critical
point the correlation length in two point correlations diverge.
Increasing $U$ the system behaves as thermal (with infinite
temperature).
In the following we give a picture for the phases developed in the
model.
We have performed numerical simulations. These are exact for the single and two site
cases.  We have also made use of the Hartree-Fock-Bogoliubov (HFB)
approximation and Gutzwiller {\it ansatz} to deal with the many  body
problem.  

The rest of the paper is organized as follows.  In  section
 \ref{sec:BH}
the model and its dissipative evolution is presented.  Then, in
\ref{sec:ss},  we describe
the single site case.  A full numerical solution is compared to the
approximation used along the paper -  the Hartree-Fock-Bogoliubov (HFB) approximation.  Section \ref{sec:mb}
deals with {\it more than one} site.  We treat the dimer, where
still numerical insight is possible, for finishing with the many-body
problem within the HFB approximation.  Some conclusions are written in \ref{sec:concl}.


\section{Model and its dissipative evolution}
\label{sec:BH}
In this work we discuss stationary solutions ($\partial_t \rho^* = 0$) of
Linblad-like master equations :
\begin{equation}
\label{qme}
\partial_t \varrho = -i [H_S, \varrho]
+ \gamma \sum_j^N  L_j \varrho L_j^\dagger - \frac{1}{2} \{ L_j^\dagger
L_j, \varrho \}
\; .
\end{equation}
Here, $\varrho$ is the reduced density matrix,  $H_S$ is the
system Hamiltonian, the operators $L_j$  are the dissipators and $\{\,
, 
\, \}$ stands for the anticommutator. 
In this work we discuss the competition between Hamiltonian and
dissipative dynamics. For this reason, it results more convenient to
adopt the following units (which will be employed throughout this
paper) $\hbar = \gamma = 1$. 
This leads to a renormalized time scale $\tau = \gamma
t$.  Indeed, $\hbar = 1$ has already been used
in (\ref{qme}).
$N$ is the number of sites considered.  We will study in detail the
single site $N=1$ in section \ref{sec:ss} and the dimer $N=2$ for testing the approximations
in Appendix \ref{app:dimer}.  When moving to the many body ($N =10$)
periodic boundary conditions will be considered [Cf. Sect. \ref{sec:mb}]. 

A Linblad-like form, also known as Gorini-Kossakowski-Sudarshan-Lindblad
equation (to credit), is the most general Markovian evolution \cite{Rivas2011}.
Therefore, here we are interested in equilibrium solutions,
$\varrho^*$, for a many body problem 
which arises from the interplay between unitary (governed by $H_S$) and non-unitary dynamics,  the latter within the Markovian theory.

An evolution like (\ref{qme}) can be derived from a system-bath
Hamiltonian.  In this approach, the system, with Hamiltonian $H_S$, is
surrounded by a bath ($H_{\rm b}$) formed by a continuum set of modes.
  Both system and bath are coupled yielding $H =
H_S + H_{\rm b} + H_{\rm I}$, with $H_{\rm I}$ the interaction
Hamiltonian.  After tracing out the bath modes and assuming weak
coupling, the dynamics for the reduced density matrix $\varrho$ is
given by (\ref{qme}).  Weak coupling
  regime means that the dynamics is governed by the
system Hamiltonian, the coupling to the bath being a perturbation. The
weak coupling limit is well justified whenever the bath correlation
functions decay sufficiently fast \citep{Rivas2011}. Although these
conditions seem to be restrictive enough, equations as (\ref{qme}) are
justified and used in a lot of cases of interest.

 When one faces  such a situation, typically,  the
dissipators are such that the stationary state coincides with 
Gibbs $\rho^* \sim {\rm e}^{-\beta H}$ \cite[Sect 3.2.2]{Breuer2002}.  
This is a nice property connecting non-equilibrium
dynamics with {\it standard} thermal physics. Exceptions to the latter
come whenever the coupling can not be considered weak
\cite{Pachon2014} or by deforming
the coupling via, {\it e.g.}, driving.  If the system-bath coupling
leaves the Markovian-weak limit  the evolution is in general
much more complicated than (\ref{qme}) \cite{Breuer2002}.  However, it turns out that
via the inclusion of driving fields and ancillary systems the system-bath can still be in
this weak limit but some dissipator engineering is allowed.  
This is the case that we are going to discuss here.  We will still
assume a Linblad form but the dissipators are going to be, say {\it non
  thermal}, {\it i.e.} such that $\varrho^* \neq  {\rm e}^{-\beta H}$.

\subsection{Bose Hubbard in a squeezed dissipator}

We study the one-dimensional BH model, 
\begin{equation}
\label{BH}
H = \sum_j^N  \omega n_j
+
U n_j ( n_j - 1)
+
J (a_j^\dagger a_{j+1} + {\rm h.c.} )
\end{equation}
here $n_j = a_j^\dagger a_j$ with $a_j$ ($a_j^{\dagger}$) 
the annihilation (creation) bosonic operators on site $j$ ($[a_j, a_{j^\prime}^\dagger]=
\delta_{j j^\prime}$).

We concentrate in both local and linear dissipators:
\begin{equation}
\label{Ls}
L_j = a_j + \eta {\rm e}^{i \theta j} a_j^\dagger
\end{equation}

In \cite{Quijandria2013} it is shown that such dissipators can be
constructed by using qubits  as ancillary systems and driving the side-bands.
This
dissipation-like mechanism was also proven to drive free bosonic ($U=0$)
hamiltonians to a critical state \cite {Eisert2010, Quijandria2013}.
Thus, the model we present here is both physically realizable and has
its interest in many-body physics driven by dissipation.

\section{Methods}

We discuss here the methods used for solving \eqref{qme}, with
Hamiltonian \eqref{BH} and dissipators \eqref{Ls}.

\subsection{Hartree-Fock-Bogoliubov approximation}
\label{sect:HFB}

We
introduce  the HFB approximation.  
In the 
non-interacting case ($U=0$)  Eq. \eqref{qme} is easily solvable by
working with first  ($\langle a_i \rangle $) and second moments ($\langle
a_i a_j \rangle$, $\langle a_i^\dagger a_j\rangle$).  At $U=0$ the
average equations for the latter form a closed set.  In this limit, the
system is Gaussian.    
However, whenever $U\neq 0$ the equations for the moments form an
infinite hierarchy, coupling correlators of higher orders.
This hierarchy needs to be cut.  The
HFB approximation is a  
{\it Gaussian ansatz}. It consists on  considering  the cumulant
expansion up to second order. 
As argued, this is exact if $U=0$. 
 This approximation has been tested in a variety of
situations as you can read in Refs. \cite{Kohler2002, Rey2004,
  Holland2001, Proukakis1998, Griffin1996, Takayoshi2010}. We show
below, section \ref{sec:results},  that the  HFB  approximation is sufficient for describing
the main phenomenology.

Within the Gaussian {\it ansatz}, averages can be computed invoking 
Wick's theorem.  For our purposes, it is sufficient to consider the
 formula:
\begin{align}
\nonumber
\langle X_1 X_2 X_3 X_4 \rangle = \, &\sigma_{12} \sigma_{34} +
\sigma_{13} \sigma_{24} + \sigma_{14} \sigma_{23} 
\\  \label{Wick}
 &- 2 \langle X_1 \rangle \langle X_2 \rangle \langle X_3 \rangle \langle X_4 \rangle
\end{align}
where $\sigma_{ij} = \langle X_i X_j \rangle$. 
Writing these higher order correlators as a function of first and second order
ones, permits to find a closed set of equations.
Some algebra yields the equations for the Linbladian
\eqref{qme} with (\ref{BH}) and (\ref{Ls}):
\begin{align}
\nonumber
\partial_t \langle a_i \rangle =&
-i \Big  ( \omega 
+ 4 U ( \langle n_i \rangle - |\langle a_i \rangle|^2 ) \Big  )
\langle a_i \rangle
- i 2 U \langle a_i ^2 \rangle \langle a_i \rangle^*
\\ \nonumber
&-i J 
(
\langle  a_{i-1} \rangle 
+
\langle  a_{i+1} \rangle 
)
\\ 
&-
\frac{1}{2} ( 1 - \eta^2) 
\langle a_i \rangle
\; .
\label{1st}
\end{align}
for the first moments.  
For the second ones we introduce some notation to alleviate the
final expressions.
We define $X_{ij} := \langle a_i^{\dagger} a_j \rangle$ and
$Y_{ij} := \langle a_i a_j \rangle$, finding that,
\begin{align}
\partial_t X_{ij} = \; &  iU( Y_{ii}^{*} Y_{ij} + 2 X_{ij} X_{ii} -
Y_{ij}^{*} Y_{jj} -2 X_{ij} X_{jj}
\nonumber \\
& -2 \langle a_i \rangle^{*2} \langle a_i \rangle \langle a_j \rangle + 2 \langle a_i \rangle^* \langle a_j \rangle^* \langle a_j \rangle^2 ) 
\nonumber\\ 
&+i J (X_{i-1,j} - X_{i,j+1} + X_{i+1,j} - X_{i,j-1})
 \nonumber\\
&- X_{ij} + \eta^2 X_{ij} + \eta^2 \delta_{ij} 
\; ,
\label{Xlm}
\end{align}
and,
\begin{align}
\partial_t Y_{ij} = & -2i \omega Y_{ij} 
\nonumber \\
&- iU (2 X_{ji} Y_{jj} +
4 X_{jj} Y_{ij} + 2 X_{ij} Y_{ii} 
\nonumber\\  
&  + 4 X_{ii} Y_{ij} + 2 \delta_{ij} Y_{jj} + 2 \delta_{ij} Y_{ii} 
\nonumber\\ &-4 \langle a_j \rangle^{*} \langle a_i \rangle \langle a_j \rangle^2 - 4 \langle a_i \rangle^{*} \langle a_j \rangle \langle a_i \rangle^2 ) 
\nonumber \\ 
&-iJ (Y_{i,j+1} + Y_{j,i+1} + Y_{i,j-1} 
 + Y_{j,i-1}) 
\nonumber \\
&- Y_{ij} - \eta {\rm e}^{i \theta j} \delta_{ij} + \eta^2 Y_{ij}  
\label{Ylm}
\end{align} 
With these equations at hand it is possible to solve the nonlinear set
of $N\times N$
 equations numerically for a reasonably large $N$.

\subsection{Gutzwiller {\it ansatz}}
\label{sec:Gu}

The Gutzwiller {\it ansatz} imposes a factorized form for the density
matrix:
\begin{equation}
\label{rhoG}
\varrho = \otimes_ j^N \varrho_j
\end{equation}
  Assuming translational invariance ($N\to \infty$ or periodic
  boundary conditions), the problem is
  reduced to a single site, non-linear master equation that can be
  numerically solved, imposing a cuttoff in the Fock space dimension.

 The dynamics within the factorized form, (\ref{rhoG})  is easy to obtain noticing that ${\rm tr} (\varrho_j)=1$
  (${\rm  tr} ( \partial_t \rho_j ) = 0$).  The final expression is:
\begin{align}
\nonumber
\partial \varrho_j =& -i [\omega n_j + U n_j (n_j -1) + J \langle a_j
\rangle a_j^\dagger + {\rm h.c.}, \rho_j]
\\ 
&+
L_j \rho_j L_j^\dagger - \frac{1}{2}  \{L_j^\dagger L_j, \varrho \}
\end{align}
with $\langle a_j \rangle \equiv {\rm tr} ( a_j \varrho )$.  Writing
writing a set of equations for the density matrix elements
$[\varrho_j]_{nm}$ we obtain a nonlinear set of equations.
  We solve the time evolution for  $[\varrho_j]_{nm}$.  In the long
  time dynamics the stationary solution is found.

The factorized {\it ansatz}, Eq.  \eqref{rhoG}, catches short
distance correlated states.  
However, the interacting (local) part $U n_j (n_j-1)$ is fully taken
into account. In this sense, Gutzwiller
is complementary to the HFB approximation.

\subsection{Numerical solution}
\label{sec:num}

These two approximations will be corroborated with {\it exact} numerical  solutions. 
Notice that for one or two sites ($N=1,2$)  the Linblad evolution can be solved
numerically.  In this paper we have  performed numerical solutions using the quantum
optics toolbox for MATLAB \cite{Tan2002}. The truncation of the Fock
space dimension, with a good degree of confidence, follows from the
comparison of numerical results with exact analytical predictions for
the non-interacting model ($U=0$) [Cf. Fig. \ref{fig:1site-GN} (blue lines)]. 

\section{Results}
\label{sec:results}

\subsection{Non interacting case}
\label{sec:nic}

The limit $U=0$ was studied in \cite{Quijandria2013}.  In a nutshell, 
dissipation-induced critical behaviour was found there. In
momentum-space, the role of the Linblad operators in the QME was to
entangle pairs of modes whose sum of momenta was equal to the driving
phase $\theta$.
Writing (\ref{qme}) in momentum space yields ($a_k = N^{-1} \sum_j
{\rm e}^{-i j k} a_j$),
\begin{equation}\label{QMEqui}
d_t \varrho = 
\sum_k -i \omega_k [a_k^\dagger a_k , \varrho] + 
\Gamma ( 2 b_k \varrho b_k^\dagger - \{b_k^\dagger b_k, \varrho\}) 
\end{equation}
with 
\begin{equation}
b_k = a_k + \eta a^\dagger_{-k+q}
\end{equation}
{\it i.e.} the modes $b_k$ are two mode squeezed operators, $q=
\theta$ and $\omega_k = \omega + 2 J cos(k)$ being the normal frequencies. 

Before going on with the discussion, we would like to introduce a quantitative definition of the {\it quadrature squeezing}. For a $N$-mode system with annihilation operators $a_j$, $j=1, ... , N$, the corresponding Hermitian quadrature operators are defined as follows 

\begin{equation}
X_i = \frac{1}{\sqrt{2}} (a_i^{\dagger} + a_i)
\end{equation}  

\begin{equation}
P_i = \frac{i}{\sqrt{2}} (a_i^{\dagger} - a_i)
\end{equation}  

Squeezing involves the second order moments of the quadrature operators. These in turn define the covariance matrix $\gamma$ 

\begin{equation}\label{covariance}
\gamma_{ij} = \frac{1}{2} \langle R_i R_j + R_j R_i \rangle - \langle R_i \rangle \langle R_j \rangle
\end{equation}

with $R = (X_1, P_1, X_2, P_2, ..., X_N, P_N)$ (or alternatively $R = (X_1, X_2, ..., X_N, P_1, P_2, ..., P_N)$. In this work we have chosen the first convention). Following \cite{Simon0} we formulate the squeezing criterion as follows: {\it a multimode system is said to be squeezed whenever the smallest eigenvalue of its covariance matrix is smaller than $1/2$}. We should point out that the ``size" of the minimum eigenvalue and the squeezing are complementary quantities. Having a big amount of squeezing implies that the minimum eigenvalue is very small ($\ll 1/2$). For example, when we say that there is an infinite amount of squeezing, we refer to the limiting situation in which the smallest eigenvalue of the covariance matrix approaches to zero.

By looking at the master equation (\ref{QMEqui}) and with a {\it  correct} choice of the system parameters: $\omega_k +
\omega_{-k+q}= 0$ we readily see that these two modes ($k, -k+q$) become maximally entangled.
For the rest,  a limiting case can be described.
Whenever $\omega_k, \omega_{-k+q} \gg \Gamma$ we can
perform the Rotating Wave Approximation for 
the dissipators and the modes will reach a thermal state, $\varrho
\sim {\rm e}^{-\beta^* a_k^\dagger a_k}$ with an effective
temperature, $\beta^*$ given by (\ref{eff_temp}).
The above argument will be elaborated through this paper in the more
general case of $U\neq 0$.  See next sections \ref{sec:ss} (for the
single site) and \ref{sec:mb} for the many body problem.


\subsection{Single site case: Transition to a thermal state}
\label{sec:ss}


\begin{figure}
\includegraphics[width=1.\columnwidth]{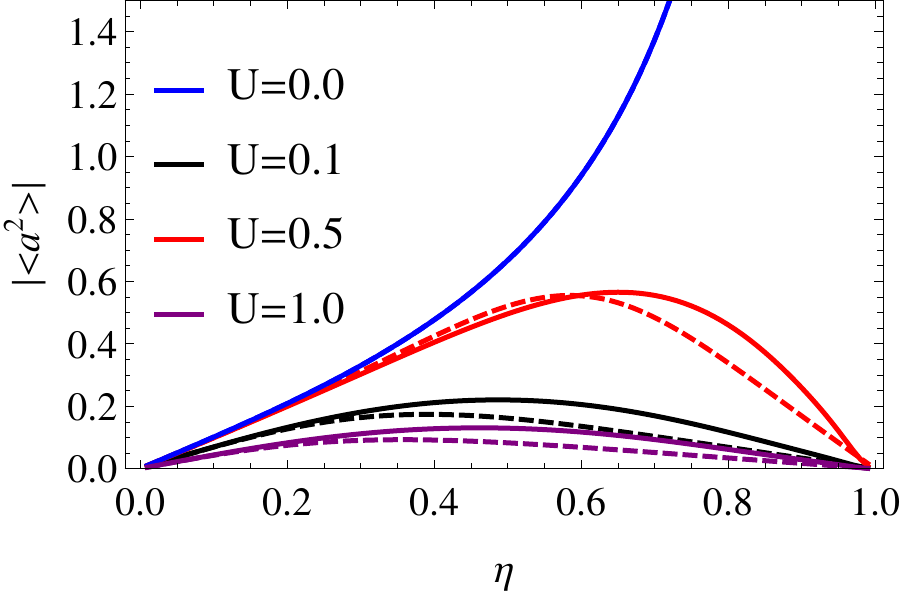}
 \caption{{\bf Single site case.} Absolute value
   of $\langle a^2 \rangle$ as a function of $\eta$ for different
   values of $U$.
   We compare the numerical solution (solid line) and the HFB
   approximation (dashed line). The numerical solution is done by
   using $N_C = 40$, where $N_C$ is the maximum number of Fock states
 considered.  The rest of parameters
are $\Gamma=1$, $\omega=0$ and $\theta=\pi/2$.}
 \label{fig:1site-GN}
 \end{figure}

Let us move to the interacting case.  We start with the single site case.
We anticipate here the main result, which is exportable to the
many-body part.
There is a competition between the photon-photon interaction, with
strength  $U$,
and the squeezed dissipators $L_j$. In the limit: $\omega, U \ll 1$,
$\varrho^*$  relaxes to a
squeezed vacuum state. On the other
hand, if $\omega, U \gg 1$, then 
$\varrho^* \sim \prod_j {\rm e}^{-\beta^* (\omega n_j + U n_j(n_j - 1) )}$ with $\beta^* = 1 /
T^*$, an
effective temperature (to be defined below).
This trade off explains the equilibrium statistics of the model
(\ref{qme}) with 
(\ref{BH}) and (\ref{Ls}). 
Importangly enough the parity symmetry $a_j \to -a_j$ is not broken,
finding always that $\langle a_j \rangle =0$.
Let us  check 
this picture.

Making $H_S=0$,  the evolution (\ref{qme}) is given by $\partial_{\tau} \varrho
= L \varrho L^\dagger - 1/2 \{L L^\dagger , \varrho \}$, with $L = a +
\eta {\rm e}^{i \phi} a^\dagger$ (for the single site).  Therefore $\varrho^* = |\xi\rangle
\langle \xi |$, with $L |\xi\rangle =0$, {\it i.e.} the  vacuum
squeezed state.
On the other hand, if
 $\omega, U \gg 1$, it is convenient to work in the
interaction picture (with respect to the $H_S$).  We have that, 
\begin{equation}
\label{rotated}
V a V^\dagger = \sum_n {\rm e}^{-i (\omega + 2 n U) t} | n \rangle
\langle n | a
\end{equation}
with  $V = {\rm exp}[i(\omega n + Un(n-1))t]$,
{\it i.e.}, the Hamiltonian rotates each Fock state with a
different phase.
Using a Rotating Wave-like argument we can expect that the time-dependent terms average to
zero.   Conserving
only the non-rotating terms we have that the quantum master equation (QME) (\ref{qme}) can be approximated by:
\begin{equation}
\label{thermal}
\dfrac{d\varrho}{d\tau} =  \frac{1}{2} (2 a \varrho a^{\dagger} -
\lbrace a^{\dagger} a ,\varrho \rbrace)  + \frac{\eta^2}{2} (2
a^{\dagger} \varrho a - \lbrace a a^{\dagger} ,\varrho \rbrace)
\end{equation}
Identifying $\eta^2 = \bar n (\omega)/(1 + \bar n(\omega))$, with
$\overline{n}(\omega)$ the Bose distribution $
\overline{n}(\omega) = {1}/{({\rm e}^{\beta \hbar \omega} - 1)} $, the
above matches the dissipators for a damped harmonic oscillator in a 
thermal bath with effective temperature,
\begin{equation}\label{eff_temp}
\beta^*  \omega = -2 \ln \eta
\; .
\end{equation}

The above argument can be validated and refined. 
It is not hard to realize that,
independently of the value of $U$, we have that:
\begin{equation}\label{n-single}
\langle n \rangle_{\varrho^*} 
\equiv \langle a^\dagger a \rangle_{\varrho^*} = \frac{\eta^2}{1 - \eta^2}
\end{equation}
Through the text we use the notation $\langle \cdot \rangle_{\varrho^*}
\equiv {\rm Tr} (\cdot \,  \varrho^*)$. 
Notice that by using (\ref{eff_temp}) in  (\ref{n-single}) we obtain
the thermal Bose distribution $\langle n \rangle_{\rho^*}  = ( {\rm
  e}^{\beta^* \omega } -1)^{-1}$.  
Therefore, for the photon number, the state is as it would be  thermal
state with the temperature predicted by the previous simple argument,
Eq. (\ref{eff_temp}).

In obtaining a dynamical equation for other
variances, as $\langle a \rangle$ and $\langle a^2 \rangle$ we find an infinity hierarchy
of equations involving higher order averages as $\langle  (a^\dagger )^n
a^m \rangle$.   We made use of the HFB or Gaussian approximation, as explained in
section \ref{sect:HFB}. 
The HFB can be justified {\it a priori} as follows.
We expect to obtain the Gaussian thermal state
$\varrho^* \sim {\rm e}^{-\beta^* a^\dagger a }$ by increasing $U$.
On the other hand, whenever $U=0$, the HFB is exact.

Particularizing Eqs. \eqref{1st} and \eqref{Ylm} to the
single site case  we can write a system of differential equations for $\langle a \rangle$,
\begin {align}
\nonumber 
\dfrac{{\rm d}\langle a \rangle}{{\rm d} t} 
=& \left( -i\omega -
  \frac{1}{2}(1 -\eta^2) \right)\langle a \rangle 
\\ 
& - 2iU 
\Big (
  2\langle n \rangle \langle a \rangle 
+ \langle a^2 \rangle \langle a \rangle^* - 2 \langle a
  \rangle^* \langle a \rangle^2  \Big )
\end {align}
and  $\langle a^2 \rangle$
\begin{align}
\nonumber
\dfrac{{\rm d}\langle a^2 \rangle}{{\rm d}t} =& \left( -i(2\omega + 2U
  + 12U \langle n \rangle) - (1 - \eta^2) \right) \langle a^2 \rangle 
\\
&+ 8iU \langle a \rangle^* \langle a \rangle^3 - \eta {\rm e}^{i \theta}
\end{align}
This, together with (\ref{n-single}), can be solved for its
steady-state. 

Apart from the aforementioned transition to a thermal state, the other key result in
this paper is the following.  We  always find that (See Appendix \ref{app:a}
for technical details):
\begin{equation}\label{a-single}
\langle a \rangle_{\rho^*} = 0
\; .
\end{equation}
Therefore, for the single site case and within the HFB, there is not a
breaking symmetry state.  Recall that the Hamiltonian \eqref{BH}
together with the dissipators \eqref{Ls} have the parity symmetry $a_i
\to -a_i$.   Further discussion will be given in \ref{sec:mb}.

The steady-state solution for $\langle a^2 \rangle$ is given by
\begin{equation}\label{a2mean}
\langle a^2 \rangle_{\varrho ^*} = \frac{-\eta {\rm e}^{i
    \theta}}{(1-\eta^2) + i 2[U (6\langle n \rangle_{\varrho ^*} + 1)
  + \omega]}
\; .
\end{equation}
We see that $\langle a^2 \rangle_{\varrho^*}$ approaches to
zero as $U \gg 1$, while $\langle a \rangle_{\rho^*}$ and
$\langle n\rangle_{\varrho^*}$ always equal their thermal averages
[Cf. Eqs. (\ref{n-single}) and (\ref{a-single})]. Therefore, the
Gaussian approximation, in the limit  $U \gg 1$, matches the
thermal state $\varrho^* \sim {\rm e}^{-\beta^* a^\dagger a }$, as expected.

To validate all this, we perform numerical solutions, as explained in
\ref{sec:num}.
In figure \ref{fig:1site-GN} we show, first, that the HFB captures well
the numerical result.  Besides, we observe that the squeezing
grows with $\eta$ whenever $U=0$ \cite{Quijandria2013}.  As soon as $U > 0$
the state approaches a thermal state with temperature $\beta ^* \sim -
\log\eta$ [Cf. Eq. (\ref{eff_temp})].  Therefore, $\eta$ favours both
  squeezing ($U=0$) and high-T thermal states ($U\neq 0$). From this
  trade-off the maximum for $\langle a ^2 \rangle_{\varrho^*}$ in
  figure \ref{fig:1site-GN} is understood.


\subsection{Many body}
\label{sec:mb}

\begin{figure}
\includegraphics[width=.49\columnwidth]{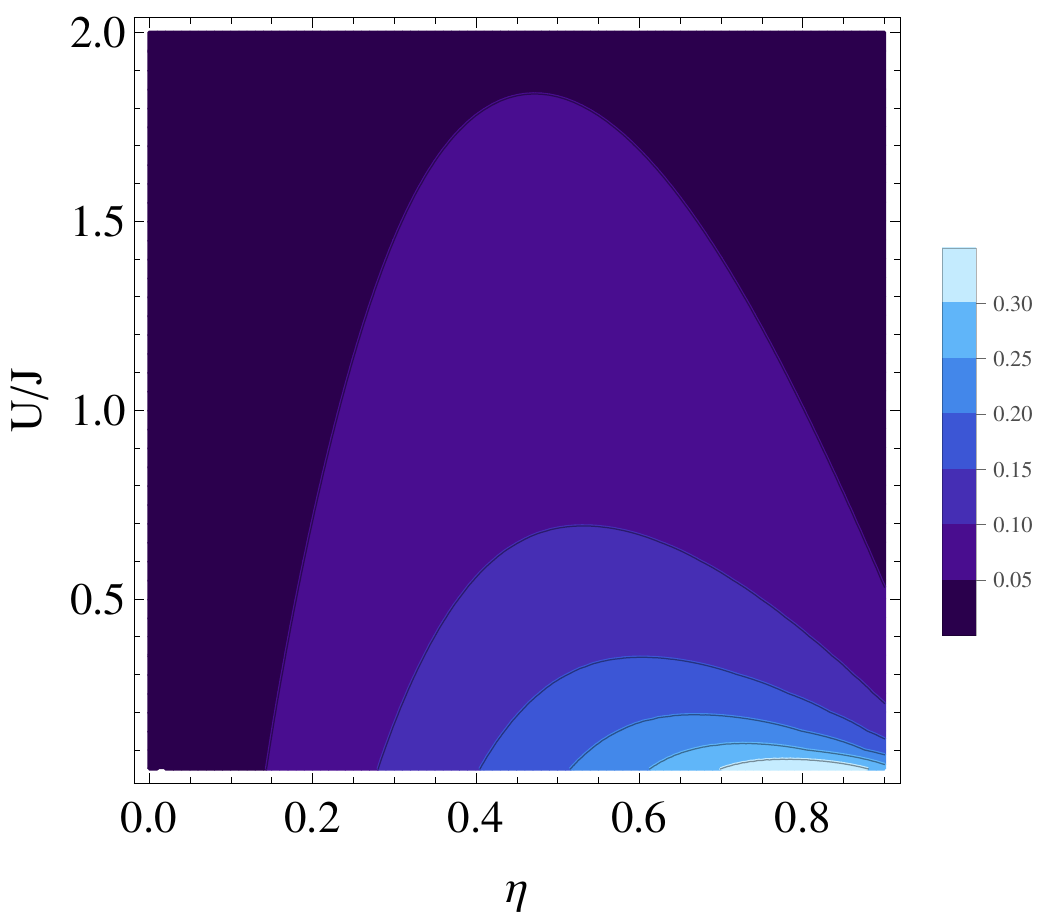}
\includegraphics[width=.49\columnwidth]{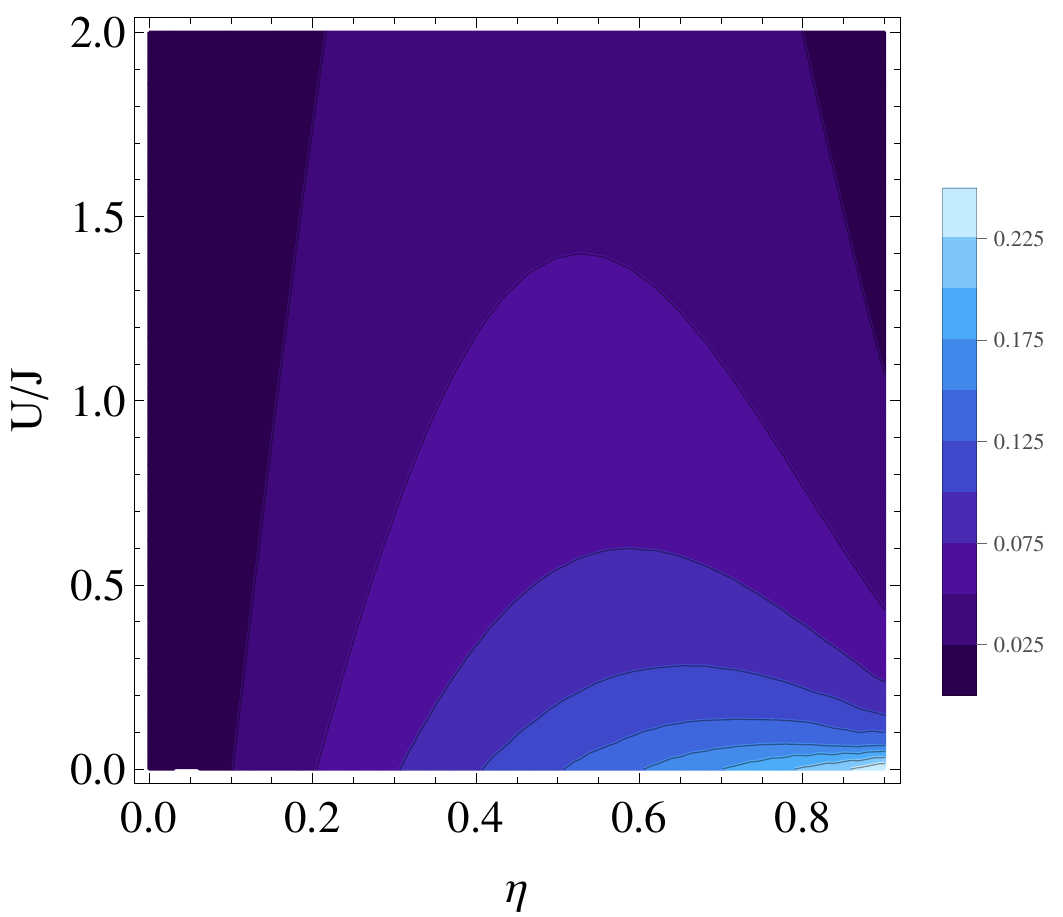}
\includegraphics[width=1.54in, height=1.2in]{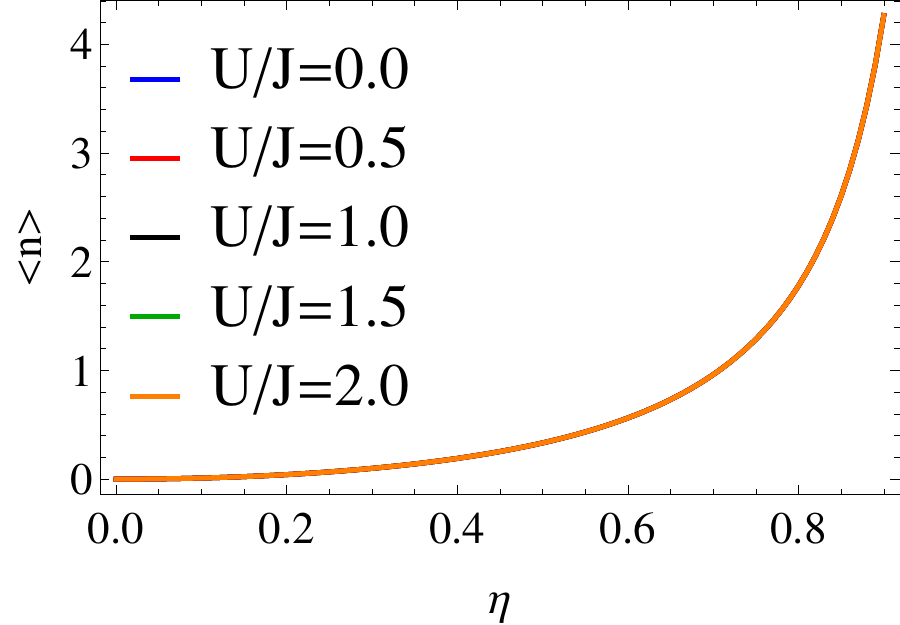}
\includegraphics[width=1.6in, height=1.2in]{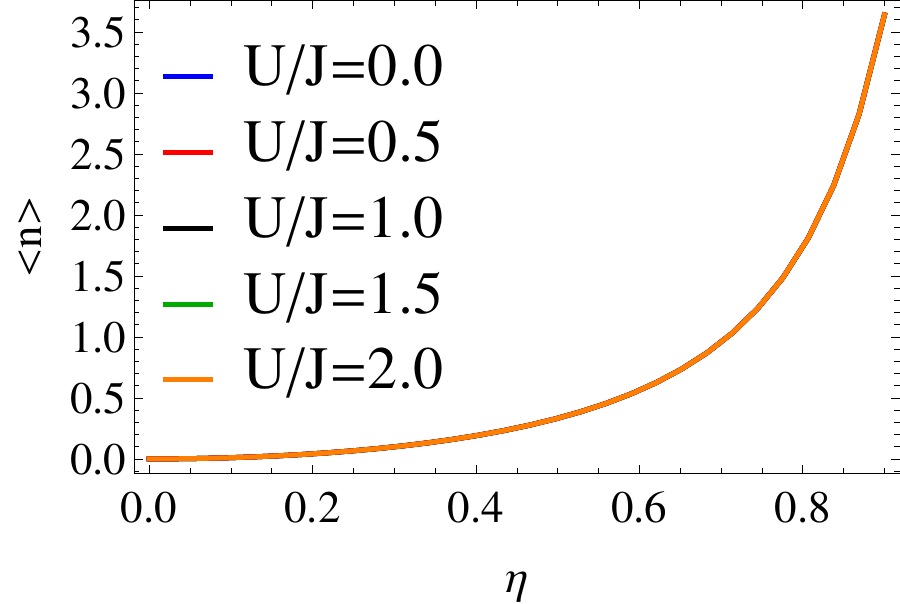}
 \caption{HFB (left) versus Gutzwiller (right). TOP: $\vert \langle a^2 \rangle \vert$ as a function of $\eta$ and $U/J$.  BOTTOM: $\langle n \rangle$ as a function of $\eta$ for some values of $U/J$. For both, HFB and Gutzwiller, we have chosen: $\omega=-2J$ and $\theta=0$. The HFB result considers a linear array with $10$ sites and periodic boundary conditions. For the Gutzwiller solution we have taken a photon cut-off $N_C = 60$.}
 \label{fig:multisite}
 \end{figure}

Equipped with the last results, we make a step
forward and discuss the many body, {\it i.e.} more than one site.
In this case, a numerical solution becomes very costly due to the violent growth of the size of the total Hilbert space. This renders the many body problem non tractable numerically.
In turn, we  have the
Gaussian approximation which in the single site case works
reasonably well [Cf. Fig. \ref{fig:1site-GN} ].
In App. \ref{app:dimer} we also test the HFB for the two site case.
Besides, 
the HFB approximation will be complemented within a {\it Gutzwiller
  ansatz}.  Combining both approaches we will capture the main physics.

\begin{figure}[t]
\includegraphics[width=.9\columnwidth]{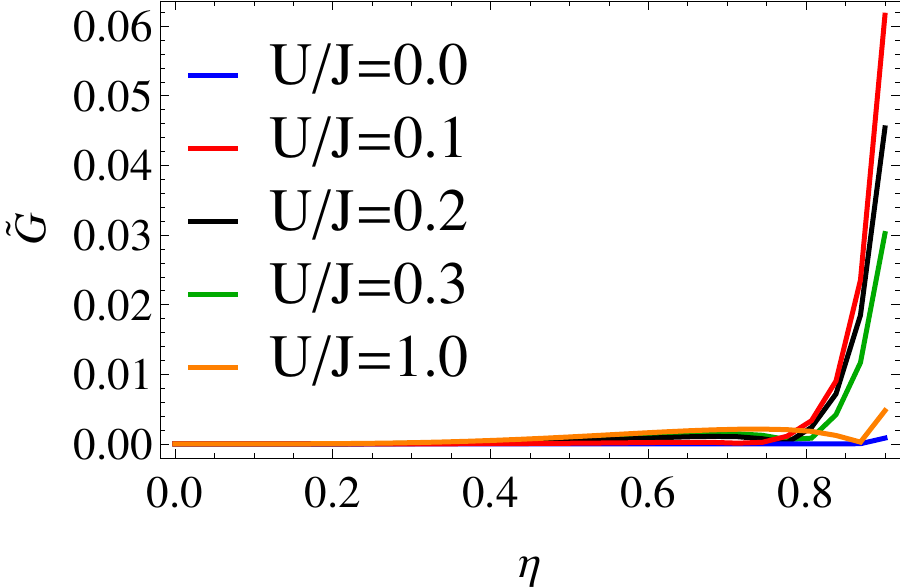}
 \caption{Non-Gaussianity $\tilde{G}$ for the Gutzwiller solution (Eq. \ref{Gdef}) as a function of $\eta$ for different values of the ratio $U/J$. Plots are shown for: $U/J=0$ (blue), $U/J=0.1$ (red), $U/J=0.2$ (black), $U/J=0.3$ (green) and $U/J=1.0$ (orange). This solution corresponds to $\omega = -2J$, $\theta = 0$ and a photon cut-off $N_C = 60$.}
 \label{fig:gauss}
 \end{figure}

We  plot in figure \ref{fig:multisite}
$|\langle a_j^2 \rangle |$ and  $\langle
a_j^\dagger a_j \rangle$,  comparing both approximations.  
 We are assuming translational invariance. 
For the HFB, systems with $N=10$ sites and periodic boundary conditions have been considered.
Thus, these quantities are
independent of $j$.
As seen in Fig. \ref{fig:multisite}  both Gutzwiller
and HFB provide essentially the same results.
We compute the {\it non-Gaussianity} for the Gutzwiller solution,
\begin{equation}\label{Gdef}
\tilde{G}:= | \langle a_j^\dagger a_j^\dagger a_j a_j \rangle 
-
\big (
2 \langle a^\dagger_j a_j \rangle^2 + \langle (a_j^\dagger)^2 \rangle
\langle a_j^2 \rangle 
- \langle a^\dagger_j \rangle^2 \langle a_j \rangle^2
\big )
|
\; ,
\end{equation}
where the last three terms come from computing the average $\langle
a_j^\dagger a_j^\dagger a_j a_j \rangle $ with the Wick formula
\eqref{Wick}, {\it i.e.} assuming a 
Gaussian distribution. A value of $\tilde{G}$ greater than zero implies that the state is non-Gaussian. In figure \ref{fig:gauss} it is clearly
appreciated that $\tilde{G}$ is always very small.  Only in a small
region for $\eta \cong 1$ and $U \cong 0.2$, $\tilde{G}$ differentiates from
zero.
\subsubsection{Transition to a thermal state}

Once the approximations have been tested, let us discuss the main
physics occurring. 
We first discuss the transition to a thermal state, pretty much
like  for  the single site [Cf. section \ref{sec:ss}].
 In the limit of large $U$, we again rotate the state  as in
 Eq. (\ref{rotated}), having that  $V a_j V^\dagger = \sum_n {\rm e}^{-i 2 n_j U t} | n_j \rangle
\langle n_j | a_j$ (in the interaction picture with respect to the self-interaction term).  The coupling $ a_j^\dagger a_{j+1} + {\rm h.c.}$ also
averages to zero within the RWA argument.  Therefore, in the limit $U$
large the effective master equation is as in (\ref{thermal}) but
summed over all the sites: $\partial_t \rho = \frac{1}{2} \sum_l (2 a_j \varrho a_j^{\dagger} -
\lbrace a_j^{\dagger} a_j ,\varrho \rbrace)  + \frac{\eta^2}{2} \sum_j(2
a_j^{\dagger} \varrho a_j - \lbrace a_j a_j^{\dagger} ,\varrho
\rbrace)$.  The stationary state, then reduces to a thermal state  of
uncoupled resonators with
temperature given by (\ref{eff_temp}). 
Further confirmation of the above picture within the HFB approximation comes from studying the $X_{ij}$ terms in the thermodynamic limit ($N \to \infty$).  Assuming translational invariance  it is easy to see that we can obtain a closed set of equations for the diagonal terms of (\ref{Xlm})
\begin{equation}
\partial_t X_{ii} = - (1 - \eta^2) X_{ii} + \eta^2
\end{equation}
which generalizes (\ref{n-single}) to the multi-site case. In a similar fashion we obtain that,
\begin{equation}
\langle a^\dagger_i a_i \rangle = \frac{1}{{\rm e}^{\beta^* \omega}- 1}
\end{equation}
We stress that the latter is independent of $U$.  This explains the
non-dependence on $U$ for $a_j^\dagger a_j$ in figure
\ref{fig:multisite}.  

\begin{figure}
\includegraphics[width=.9\columnwidth]{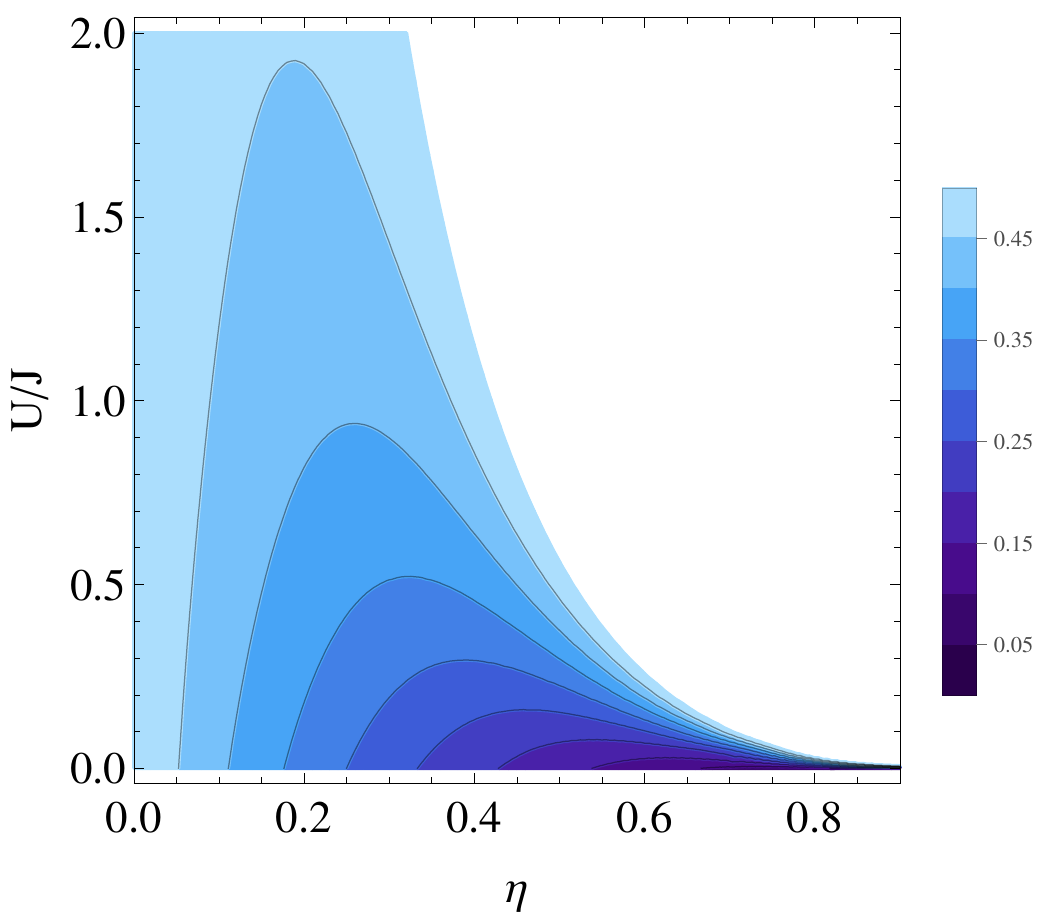}
 \caption{Squeezing (minimum eigenvalue of the covariance matrix) for a many-body array in the HFB approximation. The solution corresponds to a linear array with $10$ sites and periodic boundary conditions. The parameters chosen were $\omega = -2J$ and $\theta = 0$. The white area in the plot corresponds to no squeezing - the eigenvalues of the covariance matrix are all greater or equal than $1/2$ (according to the discussion in Sect. \ref{sec:nic}).}  
 \label{fig:sq}
 \end{figure}

The appearance of this synthetic thermal state can be traced by
computing the squeezing. For a thermal state this quadrature must equal $1/2$ (coherent state).
In figure \ref{fig:sq}  we can appreciate this transition.  To
understand it, we  must recall the non-interacting case $U=0$.  There,
the limit $\eta \to 1$ is a critical point where a couple of modes
become a maximally entangled EPR state.  In other words, the squeezing
is infinite in this point.  However as soon as $U \neq 0$ $\varrho^*$
approaches to a thermal state, with temperature given by
(\ref{eff_temp}), {\it i.e.} infinite as $\eta \to 1$.  Therefore the
squeezing becomes neglible as soon as $U\neq0$ for such a big $\eta$.
For smaller $\eta$ the thermal state has a lower temperature and the
squeezing survives for higuer $U$.

\subsubsection{No symmetry breaking }

The non dissipative BH model exhibits a $U(1)$ symmetry ($a_j
\rightarrow a_j {\rm e}^{i \phi}$). The latter is broken whenever the
expectation value of $a_j$ becomes diferent from zero (Mott insulator
- superfluid transition \cite{Sachdev}). In our case, Eq. (\ref{qme})
does not exhibit this symmetry, but it is symmetric under the parity transformation $a_j \to -a_j$.
We have found that the latter symmetry is never broken, as we always
obtain that $\langle a_j \rangle =0$.

In the non-interacting case \cite{Quijandria2013}, the parity symmetry
is not broken and $\langle a_j \rangle =0$ always holds.  
For the single site (section \ref{sec:ss}) we already learnt that this is
also the case.
We ask ourselves how this picture gets modified as soon as $U \neq 0$,
and more sites enter in the game.

In order to provide a strong argument, we are going to proceed in two directions.  First of all, 
we will follow the HFB approximation by solving the coupled equations (\ref{1st}),
(\ref{Xlm}) and (\ref{Ylm}). In second place, we will adopt a Gutzwiller {\it ansatz}.  Here,  translational
invariance will be also assumed. Even though this condition provides this ansatz of a mean field character, it is important
to stress that this approach goes beyond the HFB treatment (as already mentioned, the Gutzwiller {\it ansatz} takes fully into account the 
interaction term).
The set of parameters to investigate $(\omega/\Gamma, \theta, \eta,
J/\Gamma, U/\Gamma)$ is huge. As we have already verified, the role of the on-site potential is to thermalize the state and therefore destroy the entanglement. Thus, a very favourable set of parameters is the one which maximizes the entanglement for $U/J=0$. This is achieved by setting
 $\omega = - 2J$ and $\theta =0$ (that is, we impose that the zero
momentum mode is maximally entangled (squeezed) in the absence of interaction).
This seems reasonable due to the following argument. In the Bose-Hubbard model without dissipation, the ground state in the regime $U/J \rightarrow \infty $ is a Mott insulator with a well defined number of excitations per site, thus, $\langle a_i \rangle =0$. In the opposite limit $U/J \rightarrow 0$, the ground state is characterized by a Gutzwiller {\it ansatz} corresponding to a product state with different particle number per site \cite{Rokhsar}. Therefore, $\langle a_i \rangle \neq 0$. The latter, the superfluid phase,  corresponds to the presence of long-range correlations. Long-range ordering (divergent entanglement) in the present setup, is achieved for $U/J = 0$ and $\eta = 1$. Therefore, we could expect to find a broken symmetry around this configuration. We have
always found that $\langle a_i \rangle =0$ both in the HFB approximation 
and the Gutzwiller {\it ansatz}.  Other parameter regimes were
investigated but no symmetry breaking was found. Therefore, as we had anticipated, this model does not exhibit a phase transition.

\section{Conclusions}
\label{sec:concl}

We have studied  the equilibrium statistics of a  Bose Hubbard model
with squeezed dissipation. 
To set in a context, we mention that our model has not an external
driving competing with driving, as for example in \cite{LeBoite2013,
  Grujic2013, Boite2014}.  The
driving is, say, incoherent as introduced by the dissipators.  
In this sense, the physics discussed here has not any time dependence.
It is the squeezing, generated via the dissipators, and the
Hamiltonian competition which provides the equilibrium phases.

In summary,   we have taken as a reference the limit of zero onsite repulsion ($U=0$).
This linear model 
was shown to be critical \cite{Quijandria2013, Eisert2010}. 
In this work we have shown that as soon as $U\neq 0$ correlations shrink to
zero.
The stationary state approaches  a trivial thermal state of uncoupled
oscillators.
The temperature of this synthetic state is
proportional to the squeezing in the
dissipators,  given by Eq. (\ref{eff_temp}).
We emphasize that the dissipators (\ref{Ls}) are not
$U(1)$-symmetric, but they conserve the parity $a_j \to -a_j$.
Furthermore, it has been shown that $\langle a_j \rangle = 0$ always. 
Thus, there is no condensation.


Our findings were based on two approximations, the HFB and the
Gutzwiller {\it ansatz}.  The HFB is a Gaussian approximation [See
Sect. \ref{sect:HFB}].  The Gutzwiller assumes a factorized density
matrix as explained in \ref{sec:Gu}.  These approximations
can be understood as complementary: the HFB accounts for long distance
correlators but it is approximate in the interacting part. On the other hand,the
Gutzwiller can not catch long distance correlations but it is exact
in the nonlinearities. The physics of the problem treated
here provides an agreement between both approximations.  The
equilibrium state is basically a thermal (Gaussian) state for
uncoupled oscillators.  \\\\

\section*{Acknowledgements}

We acknowledge support from the Spanish DGICYT under
Projects No. FIS2009-10061 and No. FIS2011-25167, by the
Aragon (Grupo FENOL), QUITEMAD S2009-ESP-1594,
and the EU Project PROMISCE. The authors would also like
to  acknowledge the Centro de Ciencias de Benasque Pedro 
Pascual for its hospitality.
\\\\

\begin{figure*}
\includegraphics[width=.6\columnwidth]{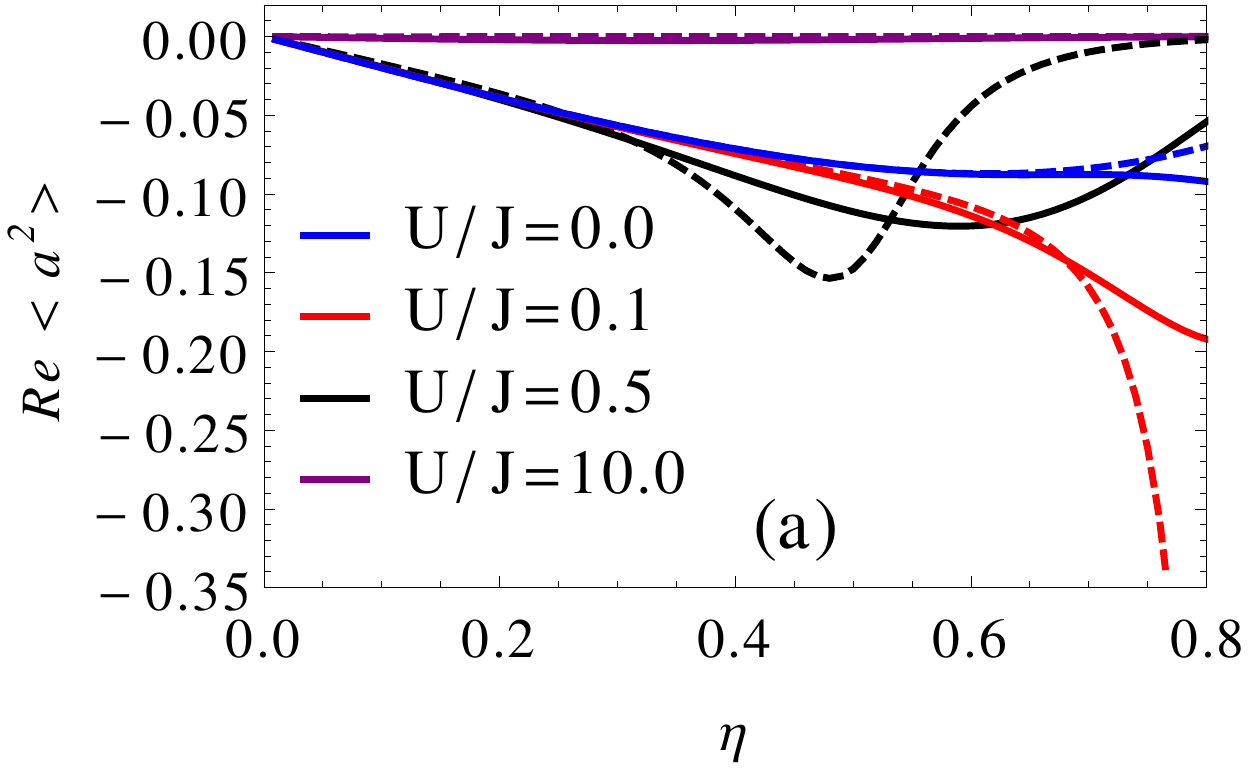}
\includegraphics[width=.6\columnwidth, height=1.28in]{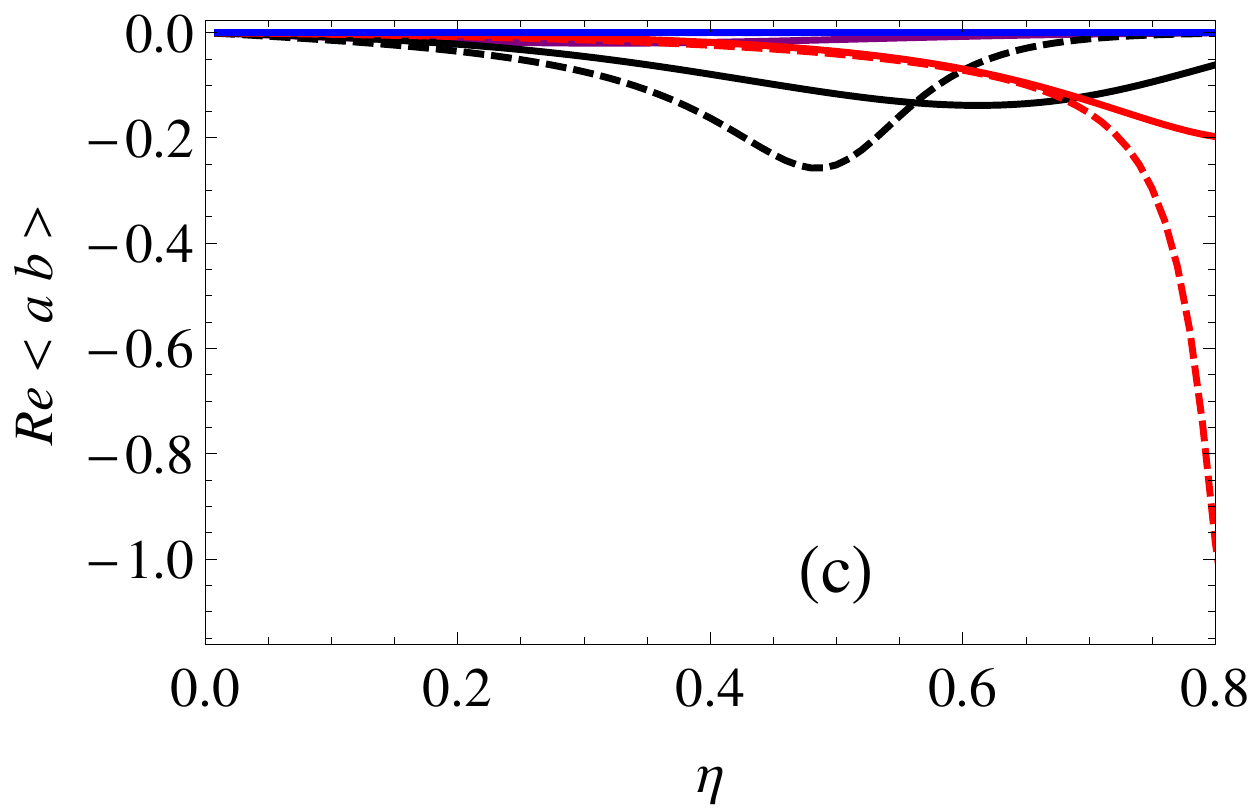}
\includegraphics[width=.6\columnwidth]{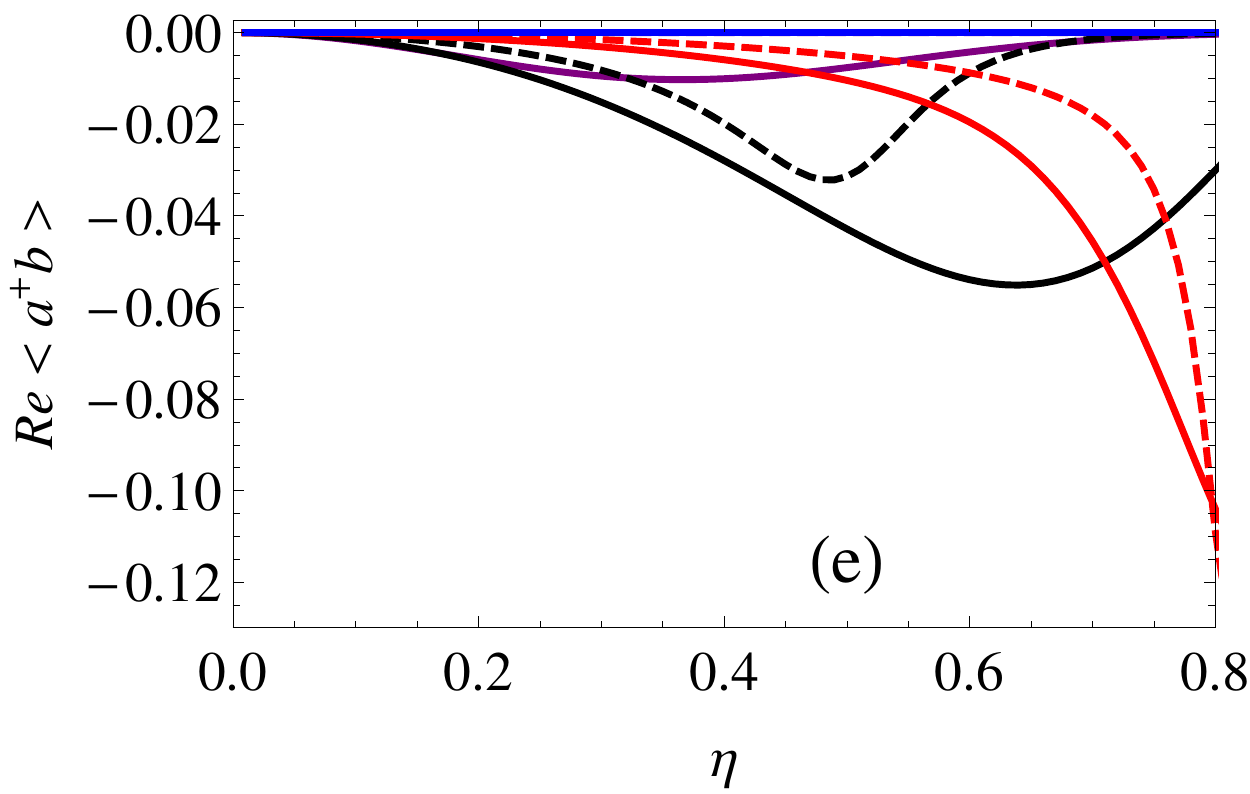}
\includegraphics[width=.6\columnwidth]{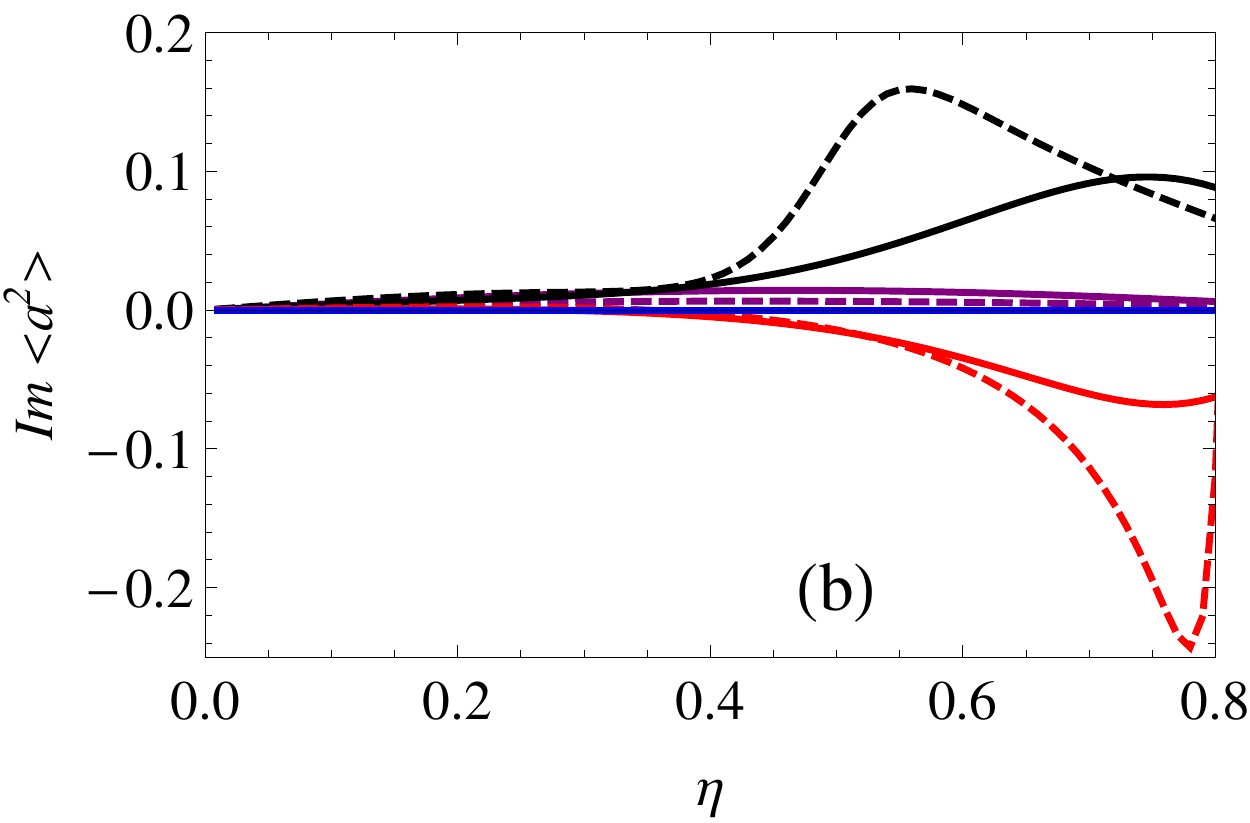}
\includegraphics[width=.6\columnwidth]{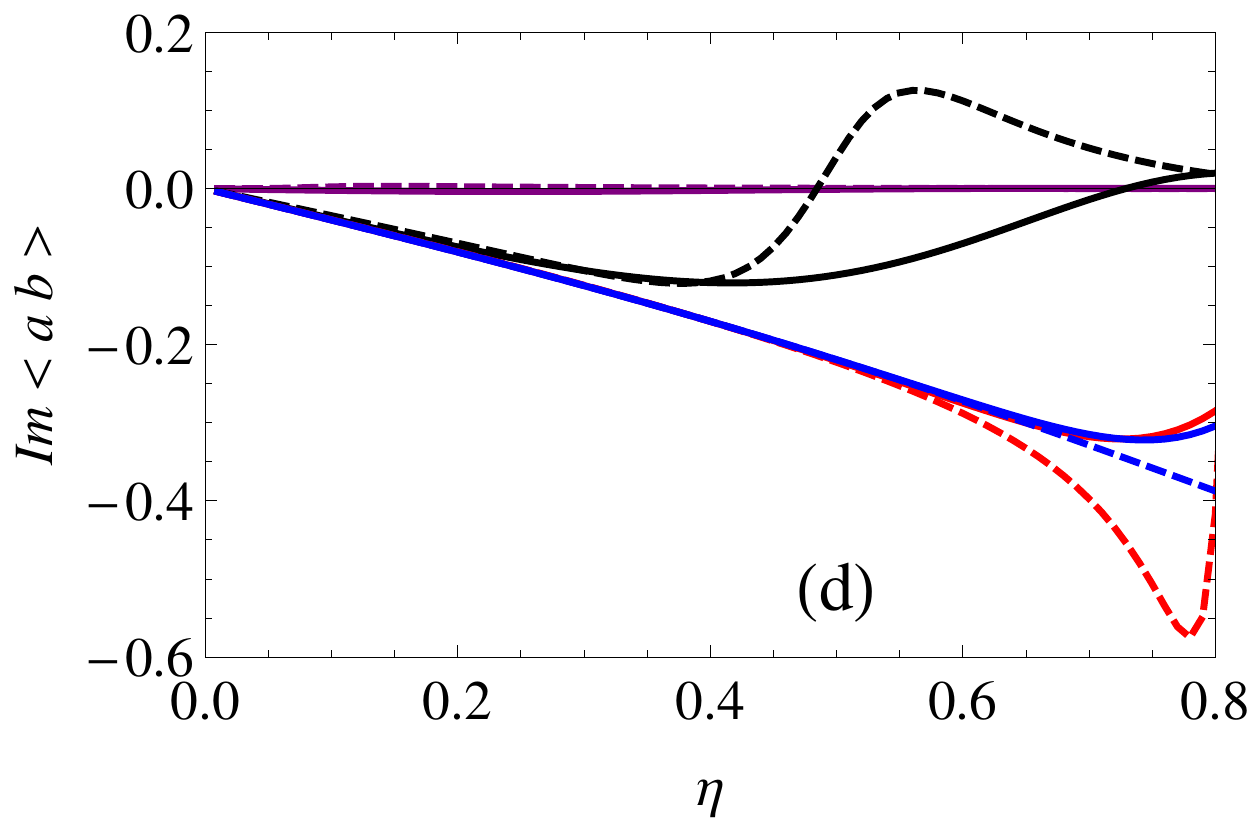}
\includegraphics[width=.6\columnwidth, height=1.32in]{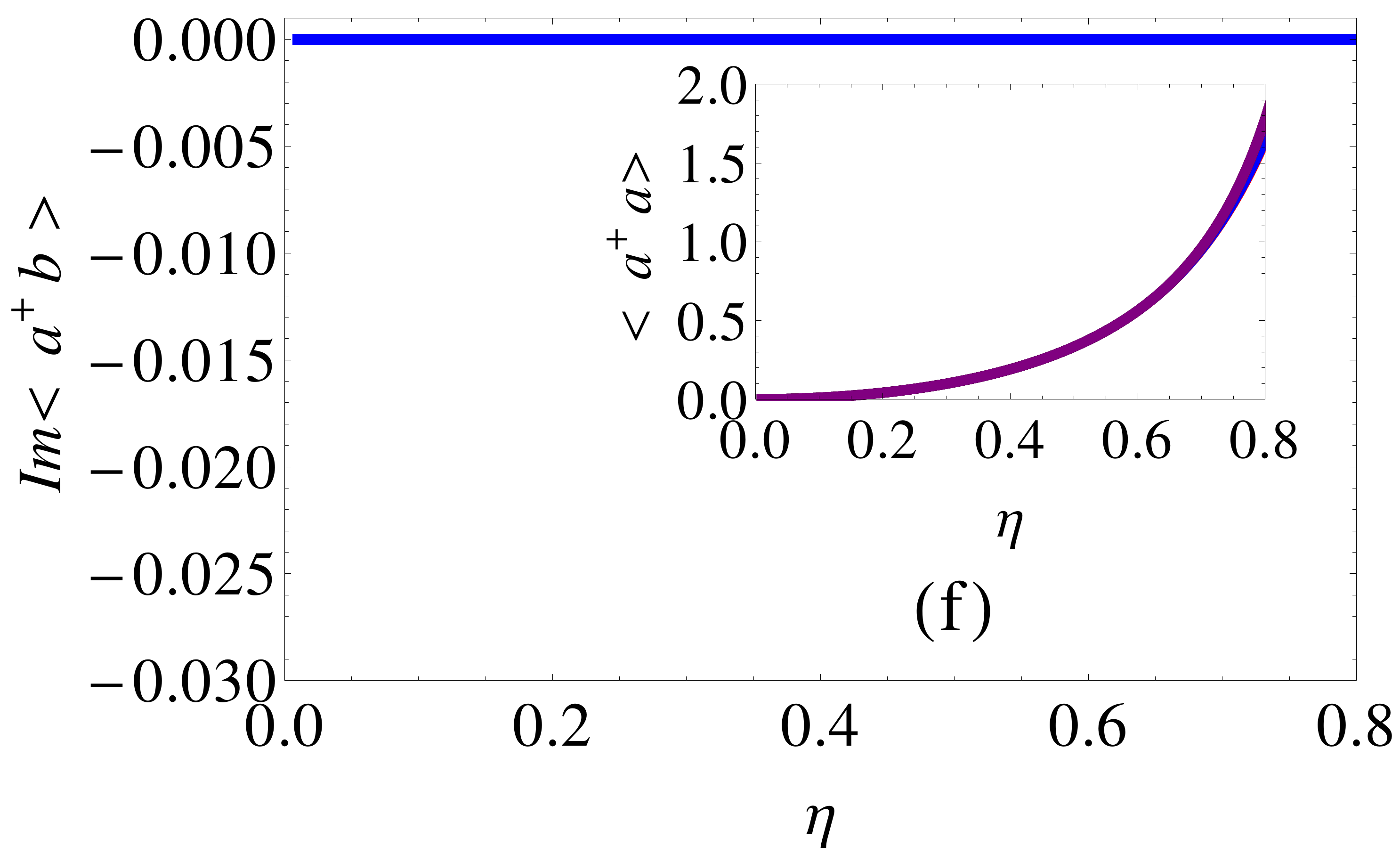}
 \caption{{\bf Two site ($N=2$) case} (a) Real and (b) imaginary part of $\langle a^2 \rangle$, (c) real and (d) imaginary part of $\langle a b \rangle$, (e) real and (f) imaginary part of $\langle a^{\dagger} b \rangle$ (Inset: number of photons in one of the cavities) all of them as a function of $\eta$ for different values of the ratio $U/J$.  We compare the full numerical solution (solid line) and the HFB
   approximation (dashed line). For both, numerical and approximate solutions, we have considered the following parameters: $\omega = 0$ and $\theta = \pi$ (this choice maximizes the entanglement at $U=0$ in the case of having only two sites). The numerical solution have been performed with a photon cut-off of  $N_C = 20$.
}
 \label{fig:2site-GN}
 \end{figure*}
  
\begin{appendix}

\section{Solutions for $\langle a \rangle$}
\label{app:a}

We detail here our steps for checking that $\langle a_i \rangle =0$.
For the single site case it is possible to argue analytically, the equations (within the HFB approximation) are:
\begin{align}
\label{a-app}
\partial_t
\langle a  \rangle
&= 
\Big [
-i \omega
-
\frac{1}{2}
( 1 - \eta^2 ) 
+
2 i U |\langle a \rangle|^2 \Big ]
\langle a \rangle
\\ \nonumber 
&-
2 i U 
\Big (
2 \langle n \rangle \langle a \rangle + \langle a^2 \rangle \langle a
\rangle^*
\Big )
\end{align}
and,
\begin{align}
\label{aa-app}
\partial_t 
\langle a^2 \rangle
=
\Big [ &
-i ( 2 \omega + 2 U + 12 U \langle n \rangle )
-
(1 -\eta^2) \Big] \langle a^2 \rangle
\\ \nonumber
&+
8 i U |\langle a \rangle|^2
\langle a \rangle^2
- \eta {\rm e}^{i \theta}
\end{align}
with $\langle n \rangle = \eta^2 / (1 - \eta^2)$ as given by Eq. (\ref{n-single}).
This is a nonlinear set of equations, we did not known how to solve
the general case analytically.
We are interested in the equilibrium solution.  Therefore we are
searching for solutions $\partial_t \langle a  \rangle = 0$ and
$\partial_t \langle a^2 \rangle = 0$.

We realize that $\langle a \rangle =0$ is always a solution of the
system, indeed for $U=0$ it is the only solution.  We want to check if $\langle a \rangle \neq 0$ is also
solution. Assuming continuity, we suppose that for $U \neq 0$ exists $\langle a \rangle = \epsilon$ with
$| \epsilon| << 1$. Then, we linearize (\ref{a-app}) and (\ref{aa-app}) discarding the terms
with $|\langle a \rangle|^2$.
Proceeding in this way, (\ref{aa-app}) becomes a closed equation for
$\langle a^2 \rangle$ with solution  given by (\ref{a2mean}). 
Formula (\ref{a2mean}) is introduced in (\ref{a-app}) obtaining a
linear set for both the real and imaginary parts of $\langle a
\rangle$:
\begin{widetext}
\begin{equation}
\left (
\begin{array}{cc}
2 U {\rm Im} [\langle a^2 \rangle] - \frac{1 -\eta^2}{2}
&
4 U \langle n \rangle - 2 U  {\rm Re} [\langle a^2 \rangle]
\\
- 4 U \langle n \rangle - 2 U {\rm Re} [\langle a^2 \rangle]
&
-2 U {\rm Im} [\langle a^2 \rangle] - \frac{1 -\eta^2}{2}
\end{array}
\right )
\;
\left (
\begin{array}{c}
{\rm Re} [\langle a \rangle]
\\
{\rm Im} [\langle a \rangle]
\end{array}
\right )
= 
0
\end{equation}
\end{widetext}
In our search for a non-trivial solution, we force the determinant
of the above matrix to be zero obtaining the condition:
\begin{equation}
| \langle a^2 \rangle |^2 
=
4 \langle n \rangle ^2 
+
\frac{\eta^4}{ 16 U^2 \langle n \rangle^2}
\end{equation}
A graphical solution of the above shows that this condition never
holds.  Indeed we can see that $ 4 \langle n \rangle ^2 
+
\frac{\eta^4}{ 16 U^2 \langle n \rangle^2} > |\langle a^2 \rangle |^2$
always.

This argument was also tested numerically, searching for solutions to the full
nonlinear set of equations (\ref{a-app}) and (\ref{aa-app}).  In the
range explored $0 < U < 10$ and $0 < \eta < 0.99$ the
only solution we found was the trivial
 $\langle a \rangle = 0$. If we perform a mean field approximation to the many-body equations, {\it i.e.} replacing the hopping term by $J (\langle a_j^\dagger \rangle a_j +
{\rm h.c.})$, the problem is reduced to the single site case already discussed. The only difference is that the onsite frequencies get shifted by $\omega \rightarrow \omega + J$. Then, in mean field approximation and within the HFB approximation no broken
symmetry is expected.
When solving the full set of HFB equations for the
multisite case (\ref{1st}-\ref{Ylm}), we always confirmed that
$\langle a_i \rangle = 0$ for $0 < U < 1$ and $0 < \eta < 0.8$.


\section{Two site case: A critical analysis of the HFB approximation}
\label{app:dimer}

In this appendix we test the HFB for the dimer ($N=2$).   
 We compute the second
moments both numerically and within the HFB.  We plot the comparison
in figure (\ref{fig:2site-GN}).  Some comments are relevant.
 Appealing to 
our experience with the single site, 
the population in each
site diverges as $\eta \to 1$ in Eq. (\ref{n-single})  [See also Fig. \ref{fig:2site-GN} f].  Therefore, our numerics
fail in this limit.  Our accuracy tests do not permit  to show
results for $\eta > 0.8$.
  In this
case we observe that for high nonlinearities ($U \cong 0.5$) the HFB
is not accurate at intermediate values of $\eta$ ($0.4 < \eta  < 0.8$).  For higher values of
$\eta$ we expect things to get better (in fact, the HFB results
clearly show this behavior). 
A similar behavior was found for the single site case. This can be
understood on the grounds of the synthetic thermal state approach
developed for it.  There we observed that, for low values of the
nonlinearity and high values of $\eta$, the steady state exhibited the
behavior of a thermal state with large  temperature $\beta^* \sim -\ln \eta$.
 Increasing the value of $U$ means that $\varrho^*$
approaches a thermal state $\sim {\rm e}^{\-\beta \omega \sum
  a_j^\dagger a_j}$ which is gaussian [Cf. Sect. \ref{sec:ss}].

\end{appendix}


\bibliographystyle{apsrev4-1}
\bibliography{BH-squeezed}

\end{document}